\documentclass[aps,preprint,pre,amsmath,amssymb]{revtex4-1}
\usepackage[abs]{overpic}
\usepackage{graphicx}% Include figure files
\usepackage{dcolumn}% Align table columns on decimal point
\usepackage{bm}% bold math
\usepackage{placeins}
\usepackage{subfig}
\usepackage{color}
\usepackage[normalem]{ulem}
\begin{document}

%\preprint{APS/123-QED} 

\title{Irreversibility-inversions in 2 dimensional turbulence}% Force line breaks with \\

\author{Andrew D. Bragg}
\email{andrew.bragg@duke.edu}
\affiliation{Department of Civil and Environmental Engineering, Duke University, Durham, North Carolina, USA}
\author{Filippo De Lillo}
\author{Guido Boffetta}
\affiliation{Department of Physics and INFN, University of Torino, via P. Giuria 1, 10125 Torino, Italy}
\date{\today}% It is always \today, today,
       % but any date may be explicitly specified

\begin{abstract}

In this paper we consider a recent theoretical prediction (Bragg \emph{et al.}, Phys. Fluids \textbf{28}, 013305 (2016)) that for inertial particles in 2D turbulence, the nature of the irreversibility of the particle-pair dispersion inverts when the particle inertia exceeds a certain value. In particular, when the particle Stokes number, ${\rm St}$, is below a certain value, the forward-in-time (FIT) dispersion should be faster than the backward-in-time (BIT) dispersion, but for ${\rm St}$ above this value, this should invert so that BIT becomes faster than FIT dispersion. This non-trivial behavior arises because of the competition between two physically distinct irreversibility mechanisms that operate in different regimes of ${\rm St}$. In 3D turbulence, both mechanisms act to produce faster BIT than FIT dispersion, but in 2D turbulence, the two mechanisms have opposite effects because of the flux of energy from the small to the large scales. We supplement the qualitative argument given by Bragg \emph{et al.} (Phys. Fluids \textbf{28}, 013305 (2016)) by deriving quantitative predictions of this effect in the short time limit. We confirm the theoretical predictions using results of inertial particle dispersion in a direct numerical simulation of 2D turbulence. A more general finding of this analysis is that in turbulent flows with an inverse energy flux, inertial particles may yet exhibit a net downscale flux of kinetic energy because of their non-local in-time dynamics.
\end{abstract}

\maketitle

%\tableofcontents

%\section{First section}
%
%Article content.
%
%\subsection{A Subsection}

\section{Introduction}

The problem of particle dispersion in turbulence is important from both fundamental and practical perspectives. From a fundamental perspective, it is important because such studies relate to the Lagrangian dynamics of turbulent flows, whose study has proven to reveal deep and interesting things about turbulence \cite{falkovich01}. Indeed, it is thought that some of the central aspects of turbulence, such as the energy cascade, are really Lagrangian, not Eulerian, in nature \cite{meneveau94,wan10}. From a practical perspective, the subject is important because of its implications for problems such as pollutant dispersion, droplet mixing in clouds, and the distribution of plankton in oceans, to name but a few.

An important problem concerns the irreversibility of multi-particle dispersion in turbulence. For the case of fluid particles, irreversibility in their multi-particle dispersion is expected due to the net flux of kinetic energy among the scales of motion of the turbulence, which is the signature of irreversibility in the underlying Eulerian velocity field. Irreversibility is important not only from a theoretical perspective, but also because irreversible dispersion means that particles spread out and mix together in turbulence at different rates \cite{bragg16}, with important implications for modeling such problems. Understanding various manifestations of Lagrangian irreversibility, and its relation to irreversibility in the underlying Eulerian turbulent velocity field is something that has recently attracted considerable attention \cite{falkovich13,jucha14,xu2014,frishman2015,xu15,buaria15,buaria16,bragg16,pumir16,bragg17b}.

%Since the term irreversibility is used in many different ways in physics, it
%is worth stating explicitly what we mean by it. In the present context, by
%irreversibility we do not refer to the impossibility of a system running in
%reverse order, as is sometimes meant in statistical physics. Instead, by
%irreversibility we mean that a given process appears different forward and
%backward in time, which is the sense in which it is usually used in the fluid
%dynamics community.

In many real systems, the particles suspended in the turbulent flow are not
fluid particles (tracers), but often have inertia, are polydisperse,
non-spherical, along with many other complexities. These features can cause the
particle motion to differ in striking ways from that of fluid particles
\cite{biferale14}, and so understanding the effect of these complexities on the
way the particles disperse in turbulence is an important problem. In a recent
paper, Bragg \emph{et al.} \cite{bragg16} investigated theoretically and
computationally how inertia affects particle dispersion and its irreversibility
in turbulence. They showed that in 3D turbulence, inertia can affect the
dispersion in very profound ways, and can have a strong effect upon the
irreversibility. They argued that whereas the irreversibility of the fluid
particle dispersion arises due to fluxes in the underlying turbulent velocity
field, inertial particles experience an additional mechanism owing to their
non-local in-time dynamics. It was shown that this additional mechanism
generates inertial particle dispersion that can be much more strongly
irreversible than that of fluid particles, with the backward-in-time (BIT)
mean-square separation of the particles being up to an order of magnitude
faster than the forward-in-time (FIT) counterpart in some cases.

Due to the fact that both mechanisms lead to a faster BIT dispersion, it is not
immediately clear how to distinguish between the two effects in realistic
flows, without relying on synthetic velocity fields with ad-hoc statistical
properties. However, a suggestion comes from the result that the asymmetry in
the fluid particle dispersion, for two particles initially at separation $r$, ultimately depends on the sign of the upscale energy flux
 through scale $r$, $\mathcal{F}(r)$.  Indeed, the faster BIT separation observed in 3D turbulence is due to the fact that, according to the Kolmogorov's four-fifths law, in the
inertial range $\mathcal{F}(r)=-\langle\epsilon\rangle<0$, where
$\langle\epsilon\rangle$ is the kinetic energy dissipation rate. If one instead
considers 2D turbulence, the same law can be generalized to show that
$\mathcal{F}(r)>0$, leading to faster separation FIT than BIT, thus reversing
the asymmetry. This has in fact already been observed for fluid particles in
 \cite{faber09}. On the other hand, the irreversibility
mechanism intrinsic in the dynamics of heavy particles does not depend on the
details of the fluid flow and will still favor BIT separation. While the latter
should dominate for particles with a large inertia, thus giving faster BIT
separation, tracers and weakly inertial particles should separate faster FIT. A
transition between the two behaviors should be observed in the inertia
parameter, the Stokes number $\rm{St}$.  In the present paper we supplement
this qualitative argument for the irreversibility-inversions with a
quantitative analysis in the short time regime. 

We test the
predictions for the irreversibility-inversions using Direct Numerical
Simulations (DNS) of inertial particle dispersion in 2D turbulence.
The presence of a transition between the two behaviors would signal
the presence of two entirely distinct physical mechanisms
generating the irreversibility of inertial particle dispersion in turbulence.
In recent years, not only
has 2D turbulence been studied in depth (see \cite{boffetta12}
and references therein) but also the dynamics of inertial particles in 2D
turbulence, e.g. \cite{boffetta04,chen06,goto06,goto08,coleman09,faber09}.

Further, 2D turbulence describes behavior that is not destroyed by perturbations
in the third dimension of Quasi-2D (Q2D) turbulent flows. Such Q2D flows can
occur in nature either because of geometrical constraints on the flow or
because of imposed body-forces \cite{boffetta12}. In particular, in Q2D
turbulent flows, one can observe $\mathcal{F}(r)>0$ over a range of
$r$ \cite{musacchio2017}. 
As such, understanding inertial particle motion and the
irreversibility of their dispersion in 2D turbulence can have applications to
Q2D turbulent flows that occur in geophysical and astrophysical contexts
\cite{bracco99}. 

The outline of the rest of the paper is as follows. In \S\ref{Theory} we review
the physical mechanisms for irreversible inertial particle dispersion in
turbulence, and explain the interesting qualitative prediction that they give
rise to for 2D turbulence. In \S\ref{Theory} we also derive a new quantitative
result for dispersion in the short-time regime that supports the qualitative
predictions. Then, in \S\ref{Res} we use data from Direct Numerical Simulations
(DNS) to test the prediction and the underlying explanations.
\S\ref{Conc} is devoted to conclusions.

%%%%%%%%%%%%%%%%%%%%%%%%%%%%%%%%%%%%%%%%%%%%%%%%%%%%%%%%%%%%%%%%
\section{Theory}\label{Theory}

We consider monodisperse inertial particle-pairs subject to Stokes drag forcing
only, whose equation of relative motion is
\begin{align}
\ddot{\bm{r}}^p(t)&\equiv\dot{\bm{w}}^p(t)=\frac{1}{\tau_p}\Big(\Delta\bm{u}(\bm{x}^p(t),\bm{r}^p(t),t)-\bm{w}^p(t)\Big),
\label{peom}
\end{align}
where $\bm{x}^p(t)$ and $\bm{x}^p(t)+\bm{r}^p(t)$ are the positions of the two
particles, $\bm{w}^p(t)$ their relative velocity, and
$\Delta\bm{u}(\bm{x}^p(t),\bm{r}^p(t),t)$ is the difference in the fluid
velocity evaluated at the particle positions \cite{bragg16}. 
In this paper we shall be
interested in the case where the system is turbulent, with statistics that are
stationary, homogeneous and isotropic. As a consequence of the homogeneity,
when $\Delta\bm{u}(\bm{x}^p(t),\bm{r}^p(t),t)$ appears in statistical
expressions we shall drop the $\bm{x}^p(t)$ argument.

In the limit $\tau_p\to0$ the particles represent fluid particles whose
equation of relative motion is
\begin{align}
\dot{\bm{r}}^f(t)&\equiv\Delta\bm{u}(\bm{x}^f(t),\bm{r}^f(t),t),
\end{align}
where the superscript `$p$' has been replaced with `$f$' to denote that these
are fluid particles. The FIT and BIT mean-square separations of the particles
are denoted by $\langle\|\bm{r}^p(t)\|^2\rangle_{\bm{\xi}}$,
$\langle\|\bm{r}^p(-t)\|^2\rangle_{\bm{\xi}}$, where
$\langle\cdot\rangle_{\bm{\xi}}$ denotes an ensemble average conditioned on
$\bm{r}^p(0)=\bm{\xi}$. 
The conditioning time can be set to zero 
since we are interested in statistically stationary flows.

In Bragg \emph{et al.} \cite{bragg16} it was argued that for inertial particles, there are two distinct mechanisms 
that generate irreversible dispersion, i.e.  $\langle\|\bm{r}^p(t)\|^2\rangle_{\bm{\xi}}\neq\langle\|\bm{r}^p(-t)\|^2\rangle_{\bm{\xi}}$. Here we summarize the conceptual ideas, and refer the readers to that paper for detailed arguments. First, we define a scale-dependent Stokes number ${\rm St}_r(t)\equiv\tau_p/\tau_r(t)$, where $\tau_r(t)$ is the eddy turnover time evaluated at the scale $\|\bm{r}^p(t)\|$. Next, we note that the quantities $\langle\|\bm{r}^p(t)\|^2\rangle_{\bm{\xi}}$ and $\langle\|\bm{r}^p(-t)\|^2\rangle_{\bm{\xi}}$ are dominated by the behavior of particle-pairs that move apart and particle-pairs that move together, respectively.

When ${\rm St}_\xi\ll1$, the effect of inertia is weak, and the
dispersion irreversibility arises because the energy flux in the turbulent
field $\Delta\bm{u}$ causes particle-pairs to move together and apart at
characteristically different rates. In 3D turbulence $\mathcal{F}(r)<0$, and
this causes the particle-pairs to move together more energetically than apart,
leading to
$\langle\|\bm{r}^p(-t)\|^2\rangle_{\bm{\xi}}>\langle\|\bm{r}^p(t)\|^2\rangle_{\bm{\xi}}$.
However, in 2D turbulence $\mathcal{F}(r)<0$, thus leading to the opposite
behavior
$\langle\|\bm{r}^p(t)\|^2\rangle_{\bm{\xi}}>\langle\|\bm{r}^p(-t)\|^2\rangle_{\bm{\xi}}$.
This is referred to as the Local Irreversibility Mechanism
(LIM)~\cite{bragg16}, since it arises from the behavior of the local turbulence
experienced by the particles.

When ${\rm St}_\xi\geq\mathcal{O}(1)$, the inertial particle relative
motion is strongly affected by their non-local in-time dynamics, which gives
rise to the ``path-history effect'' when the statistics of $\Delta\bm{u}$
depend upon separation. This effect arises since particle-pair motion depends
upon the multi-scale nature of turbulence, and as inertial particles posses
memory, their motion can be influenced by their interaction with turbulent
scales in the past that had properties very different from the scales
associated with their current separation. In particular, particle-pairs moving
together will carry with them a memory of their interaction with scales larger
than those at their current separation, whereas particle-pairs moving apart
will carry with them a memory of their interaction with scales smaller than
those at their current separation. This path-history effect leads to downscale
energy fluxes in the inertial particle-pair motion \cite{bragg14b,bragg14c},
associated with inertial particles moving together more energetically than
apart, and hence leads to
$\langle\|\bm{r}^p(-t)\|^2\rangle_{\bm{\xi}}>\langle\|\bm{r}^p(t)\|^2\rangle_{\bm{\xi}}$.
In Bragg \emph{et al.} \cite{bragg16} this was refereed to as the Non-Local
Irreversibility Mechanism.

That the NLIM generates faster BIT than FIT dispersion only
depends upon the fluid having the property that the statistics of
$\Delta\bm{u}$ increase with increasing separation. Crucially, unlike the LIM,
the NLIM does not depend upon the sign of $\mathcal{F}$. This then leads to an
interesting prediction: In 2D turbulence,
$\langle\|\bm{r}^p(t)\|^2\rangle_{\bm{\xi}}>\langle\|\bm{r}^p(-t)\|^2\rangle_{\bm{\xi}}$
until ${\rm St}_{\xi}$ becomes large enough and then
$\langle\|\bm{r}^p(-t)\|^2\rangle_{\bm{\xi}}>\langle\|\bm{r}^p(t)\|^2\rangle_{\bm{\xi}}$,
i.e. an inversion in the nature of the dispersion irreversibility as the
particle inertia is increased.
%
%%%%%%%%%%%%%%%%%%%%%%%%%%%%%%%
\subsection{Short-time analysis}
The preceding qualitative argument for the irreversibility inversion may be
supplemented by a quantitative analysis of the dispersion in the short-time
limit. 
Using an expansion in $t$ we may write \cite{jucha14,buaria15} 
\begin{align}
\Big\langle\|\bm{r}^f(t)\|^2-\|\bm{r}^f(-t)\|^2\Big\rangle_{\bm{\xi}}=2\Big\langle \Delta\bm{u}(\bm{\xi},0)\bm{\cdot}\Delta\bm{a}(\bm{\xi},0)\Big\rangle t^3+\mathcal{O}(t^5),
\label{jucha}
\end{align}
where $\Delta\bm{a}(\bm{\xi},0)$ is the difference in the fluid acceleration
field evaluated at two points separated by $\bm{\xi}$. Using results from Lindborg \cite{lindborg99}, we can derive the following result for statistically stationary, isotropic, 2D turbulence 
\begin{align}
\begin{split}
\Big\langle \Delta\bm{u}(\bm{\xi},0)\bm{\cdot}\Delta\bm{a}(\bm{\xi},0)\Big\rangle&=\frac{1}{2}\bm{\nabla_\xi \cdot}\Big\langle\Delta\bm{u}(\bm{\xi},0)\|\Delta\bm{u}(\bm{\xi},0)\|^2\Big\rangle\\
&=\frac{1}{6}\Big(\xi \nabla_\xi^2 S^f_{3,\parallel}+8 \nabla_\xi S^f_{3,\parallel} \Big),
\end{split}\label{ua}
\end{align}
where $S^f_{3,\parallel}(\xi)\equiv\langle[\Delta u_\parallel(\xi,0)]^3\rangle$
and $\xi\equiv\|\bm{\xi}\|$. For forced 2D turbulence, with forcing lengthscale
$\ell_f$, we have the double cascade scenario for which \cite{boffetta12}
\begin{align}
S^f_{3,\parallel}(\xi)&=\frac{1}{8}\Omega_I \xi^3,
\quad\text{for}\,\xi\ll \ell_f,\label{dircas}\\
S^f_{3,\parallel}(\xi)&=\frac{3}{2}\varepsilon_I \xi,\,\,\,
\quad\text{for}\,\xi\gg \ell_f,
\label{invcas}
\end{align}
where $\Omega_I$ is the enstrophy injection rate and 
$\varepsilon_I \simeq \Omega_I \ell_f^2$ is the energy injection rate. 
From (\ref{dircas}) and (\ref{invcas}) we see that the contribution
in (\ref{jucha}) is always positive but with different $\xi$ dependence 
for the direct and the inverse cascade regimes. Substituting (\ref{dircas}) and (\ref{invcas}) into (\ref{ua}) gives
\begin{align}
\Big\langle\|\bm{r}^f(t)\|^2-\|\bm{r}^f(-t)\|^2\Big\rangle_{\bm{\xi}}&=\frac{5}{8}\Omega_I \xi^2 t^3+\mathcal{O}(t^5), \quad\text{for}\,\xi\ll \ell_f,\label{rfFB1}\\
\Big\langle\|\bm{r}^f(t)\|^2-\|\bm{r}^f(-t)\|^2\Big\rangle_{\bm{\xi}}&=
2 \varepsilon_I t^3+\mathcal{O}(t^5), \,\,\quad\quad\text{for}\,\xi\gg \ell_f,\label{rfFB2}
\end{align}
%
%
%\begin{align}
%\Big\langle\|\bm{r}^f(t)\|^2-\|\bm{r}^f(-t)\|^2\Big\rangle_{\bm{\xi}}=\frac{7}{8}\Omega_I \xi^2 t^3+\mathcal{O}(t^5), \quad\text{for}\,\xi\ll \ell_f,\label{rfFB}
%\end{align}
%
and these are positive, showing that FIT dispersion is faster than BIT
dispersion in both the direct and inverse cascade regimes of 2D turbulence.

For inertial particles, the result corresponding to (\ref{jucha}) is
\begin{align}
\Big\langle\|\bm{r}^p(t)\|^2-\|\bm{r}^p(-t)\|^2\Big\rangle_{\bm{\xi}}=2\Big\langle\bm{w}^p(0)\bm{\cdot}\dot{\bm{w}}^p(0)\Big\rangle_{\bm{\xi}} t^3+\mathcal{O}(t^5).\label{juchaIP}
\end{align}
To express
$\langle\bm{w}^p(0)\bm{\cdot}\dot{\bm{w}}^p(0)\rangle_{\bm{\xi}}$ in the form of a flux, analogous to  (\ref{ua}), we use the evolution equation for the PDF
$p(\bm{\xi},\bm{w},t)\equiv\langle\delta(\bm{r}^p(t)-\bm{\xi})\delta(\bm{w}^p(t)-\bm{w})\rangle$,
namely \cite{bragg14e}
\begin{align}
\partial_t p=-\bm{\nabla_\xi \cdot}p\bm{w}-\bm{\nabla_w \cdot}p\langle\dot{\bm{w}}^p(t)\rangle_{\bm{\xi},\bm{w}}.\label{PDFeqn}
\end{align}
Multiplying the stationary form of (\ref{PDFeqn}) by $\|\bm{w}\|^2$ and then
integrating over all $\bm{w}$ allows us to derive the following result for a
statistically stationary, isotropic system
\begin{align}
\Big\langle\bm{w}^p(t)\bm{\cdot}\dot{\bm{w}}^p(t)\Big\rangle_{\bm{\xi}}=\frac{1}{2g}\bm{\nabla_\xi \cdot}g\Big\langle \bm{w}^p(t)\|\bm{w}^p(t)\|^2\Big\rangle_{\bm{\xi}},\label{wwdoteqn}
\end{align}
where \cite{bragg14b}
\begin{align}
g({\xi})\equiv\frac{N(N-1)}{n^2V}\int_{\mathbb{R}^3}p\,d\bm{w},
\end{align}
is the Radial Distribution Function (RDF), $N$ is the total number of particles
in the control volume $V$, and $n\equiv N/V$. 

Using the equation of motion for ${\bm{w}^p}$ we may write
\begin{align}
\Big\langle\bm{w}^p(0)\bm{\cdot}\dot{\bm{w}}^p(0)\Big\rangle_{\bm{\xi}}=\Big\langle\Delta\bm{u}(\bm{r}^p(0),0)\bm{\cdot}\bm{w}^p(0)\Big\rangle_{\bm{\xi}}-\Big\langle\|\bm{w}^p(0)\|^2\Big\rangle_{\bm{\xi}}\label{wweqn}
\end{align}

In the regime ${\rm St}_\xi(t)\ll1$, $\bm{w}^p(t)=\Delta\bm{u}(\bm{r}^p(t),t)-{\rm St}_\xi\tau_\xi\Delta\bm{a}(\bm{r}^p(t),t)+\mathcal{O}({\rm St}_\xi^2)$, where $\Delta\bm{a}(\bm{r}^p(t),t)$ is the difference in the fluid acceleration field evaluated at the positions of the two inertial particles. In this case, we find for ${\rm St}_\xi\ll 1$
\begin{align}
\Big\langle\bm{w}^p(0)\bm{\cdot}\dot{\bm{w}}^p(0)\Big\rangle_{\bm{\xi}}=\Big\langle\Delta\bm{u}(\bm{r}^p(0),0)\bm{\cdot}\Delta\bm{a}(\bm{r}^p(0),0)\Big\rangle_{\bm{\xi}}-{\rm St}_\xi\tau_\xi\Big\langle\|\Delta\bm{a}(\bm{r}^p(0),0)\|^2  \Big\rangle_{\bm{\xi}}.\label{wweqn2}
\end{align}
The result in \eqref{wweqn2} then implies through \eqref{juchaIP} that in the regime ${\rm St}_\xi \ll1$, inertial particles will separate faster FIT than BIT in 2D turbulence, just as is the case for tracers. Note that this involves the assumption that $\langle\Delta\bm{u}(\bm{r}^p(0),0)\bm{\cdot}\Delta\bm{a}(\bm{r}^p(0),0)\rangle_{\bm{\xi}}$ differs from $\langle \Delta\bm{u}(\bm{\xi},0)\bm{\cdot}\Delta\bm{a}(\bm{\xi},0)\rangle$ in magnitude, but not in sign. This seems very reasonable, especially since preferential sampling is not too strong for ${\rm St}_\xi \ll1$. The term of order ${\rm St}_\xi$ indicates that in this regime inertia acts to reduce the asymmetry.

In the regime ${\rm St}_\xi\geq \mathcal{O}(1)$, something very different can occur.
In particular, \eqref{wweqn} shows that if
$\langle\|\bm{w}^p(0)\|^2\rangle_{\bm{\xi}}>\langle\Delta\bm{u}(\bm{r}^p(0),0)\bm{\cdot}\bm{w}^p(0)\rangle_{\bm{\xi}}$
then we can have $\Big\langle\bm{w}^p(0)\bm{\cdot}\dot{\bm{w}}^p(0)\Big\rangle_{\bm{\xi}}<0$, i.e. that although tracer particles separate faster FIT than BIT in 2D turbulence, inertial particles with ${\rm St}_\xi\geq \mathcal{O}(1)$ can separate faster BIT than FIT, an inversion in the nature of the two-particle
dispersion irreversibility. This also means through (\ref{wwdoteqn}), that the
inertial particle-pairs would experience a \emph{downscale} flux of kinetic
energy, opposite in sign to the fluid energy flux $\mathcal{F}(r)>0$ for 2D
turbulence. 

The behavior $\langle\|\bm{w}^p(0)\|^2\rangle_{\bm{\xi}}>\langle\Delta\bm{u}(\bm{r}^p(0),0)\bm{\cdot}\bm{w}^p(0)\rangle_{\bm{\xi}}$ can arise in the regime ${\rm St}_\xi\geq \mathcal{O}(1)$, at sub-integral scales (i.e. where the statistics of $\Delta\bm{u}$ depend upon separation), through the ``path-history effect'' \cite{bragg14b,ireland16a}, described earlier. Recall that this effect describes the fact that since inertial particles posses  memory, they can remember their interaction with turbulent scales along their path-history that were larger and more energetic than those at their current separation, giving rise to $\|\bm{w}^p(t)\|>\|\Delta\bm{u}(\bm{r}^p(t),t)\|$, in a statistical sense. Although the effect operates at all sub-integral scales in turbulence, it is most effective in the dissipation range (i.e. where the velocity field is smooth) where it gives rise to ``caustics'' \cite{wilkinson05}, characterized by $\|\bm{w}^p\|\gg\|\Delta\bm{u}(\bm{r}^p(t),t)\|$.

Another important mechanism influencing $\bm{w}^p(t)$ in turbulence is the
preferential sampling effect, wherein because of their inertia, inertial
particles tend to avoid vorticity dominated regions of the flow \cite{maxey87}.
At inertial range scales, preferential sampling is associated with the inertial
particles avoiding regions where the coarse-grained vorticity dominates over
the coarse-grained straining motions of the turbulence \cite{bragg14e}.
Although the energy flux $(\tau_p/2g)\bm{\nabla_\xi \cdot}g\langle
\bm{w}^p(t)\|\bm{w}^p(t)\|^2\rangle_{\bm{\xi}}$ in (\ref{wwdoteqn}) is certainly affected by
preferential sampling, it is the path-history effect, and not preferential
sampling, that should be understood as the fundamental cause of the flux
inversions described above. One argument for this is the fact that the above
discussion also applies to flows in which the temporal
evolution of $\Delta\bm{u}(\bm{r}^p(t),t)$ is white-in-time, since the
path-history effect still operates in such a flow \cite{bragg14d}. However, in
white-in-time flows, the preferential sampling effect is absent
\cite{gustavsson11b}.

In summary then, our arguments predict that although particle-pairs separate faster FIT than BIT when ${\rm St}_\xi\ll 1$, as ${\rm St}_\xi$ is increased, this behavior can invert, causing particle-pairs to separate faster BIT than FIT. This inversion occurs because when ${\rm St}_\xi$ is small, the direction of the particle-pair energy flux is governed by the flux in the turbulent velocity field, which is upscale in 2D turbulence. However, as ${\rm St}_\xi$ is increased, the non-local/path-history contribution to their dynamics becomes important, and this always causes the flux to be downscale.

Unlike the fluid particle case, we are not able in general to derive an analytical prediction for $\langle\|\bm{r}^p(t)\|^2-\|\bm{r}^p(-t)\|^2\rangle_{\bm{\xi}}$ in the short-time regime since analytical results for $(\tau_p/2g)\bm{\nabla_\xi \cdot}g\langle \bm{w}^p(t)\|\bm{w}^p(t)\|^2\rangle_{\bm{\xi}}$ are not in general available for ${\rm St}_\xi\geq\mathcal{O}(1)$. However, analytical results for the statistics of $\bm{w}^p(t)$ in the limit $\xi\to0$ and for ${\rm St}_\xi\geq\mathcal{O}(1)$ have been derived by Gustavsson \& Mehlig \cite{gustavsson11} that apply to 2D flows. Here, we make use of those results to derive a prediction for the short-time behavior of $\langle\|\bm{r}^p(t)\|^2-\|\bm{r}^p(-t)\|^2\rangle_{\bm{\xi}}$ when ${\rm St}_\xi\geq\mathcal{O}(1)$ and $\xi\ll\ell_f$.

Substituting (\ref{wwdoteqn}) into (\ref{juchaIP}), and invoking the isotropy of the system, we obtain
\begin{align}
\begin{split}
\Big\langle\|\bm{r}^p(t)\|^2-\|\bm{r}^p(-t)\|^2\Big\rangle_{\bm{\xi}}=&\Big(\Big[\nabla_\xi S^p_{3,\parallel}+\nabla_\xi S^p_{3,\parallel,\perp}+4\xi^{-1}S^p_{3,\parallel\perp}\Big] +\Big[S^p_{3,\parallel}+ S^p_{3,\parallel,\perp}\Big]\nabla_\xi\ln g\Big) t^3,
\label{juchaIP2}
\end{split}
\end{align}
where $S^p_{3,\parallel}(\xi,t)\equiv\langle[w^p_\parallel(t)]^3\rangle_{\xi}$ and $S^p_{3,\parallel\perp}(\xi,t)\equiv\langle w^p_\parallel(t)[w^p_\perp(t)]^2\rangle_{\xi}$. We will now demonstrate that when $\tau_p$ is large enough for the non-local dynamics to control the inertial particle relative velocities at $\xi\ll \ell_f$, the sign of (\ref{juchaIP2}) is negative, in contrast to (\ref{rfFB1}) which is positive. 

When the non-local dynamics control the inertial particle velocities at small separations, ``caustics'' form \cite{gustavsson11}, in which the statistics of $w^p_\parallel(t)$ and $w^p_\perp(t)$ become approximately equal \cite{ireland16a}, so that $S^p_{3,\parallel}\approx S^p_{3,\parallel\perp}$. In the caustic regime, the structure functions $S^p_{N,\parallel}(\xi,t)\equiv\langle[w^p_\parallel(t)]^N\rangle_{\xi}$ exibit power-law behavior \cite{gustavsson11} such that $S^p_{3,\parallel}=\alpha_3 \xi^{\beta_3}$, $g(\xi)=\alpha_0 \xi^{-\beta_0}$, where $\alpha_0,\beta_3,\beta_0$ are all positive (we shall return to $\alpha_3$ shortly). Substituting these results into (\ref{juchaIP2}) gives
\begin{align}
\Big\langle\|\bm{r}^p(t)\|^2-\|\bm{r}^p(-t)\|^2\Big\rangle_{\bm{\xi}}=2\alpha_3\Big(2+\beta_3-\beta_0\Big)\xi^{\beta_3-1} t^3.\label{juchaIP3}
\end{align}
The exponents take on values $\beta_3\in[0,3]$, where $\beta_3=0$ corresponds to the ballistic limit, and $\beta_0\in[0,1)$ in 2D turbulence (e.g., Boffetta \emph{et al.} \cite{boffetta04}). Consequently, $2+\beta_3-\beta_0>0$, and so the sign of (\ref{juchaIP3}) is determined by the sign of $\alpha_3$ which corresponds to the sign of $S^p_{3,\parallel}$. This is precisely what is expected given the irreversibility mechanisms explained earlier, and those arguments predict that when the non-local dynamics of the inertial particles dominate their motion then $\alpha_3 <0$, corresponding to the particle-pairs approaching more energetically than they separate. 

Taken together, the results in (\ref{rfFB1}) and (\ref{juchaIP3}) predict that in the direct cascade regime of 2D turbulence, $\langle\|\bm{r}^p(t)\|^2\rangle_{\bm{\xi}}-\langle\|\bm{r}^p(-t)\|^2\rangle_{\bm{\xi}}$ will be positive for ${\rm St}_\xi=0$, but will invert and become negative once ${\rm St}_\xi$ is large enough for the path-history effect to dominate the particle relative velocities.
%

%%%%%%%%%%%%%%%%%%%%%%%%%%%%%%%%%%%%%%%%%%%%%%%%%%%%%%%%%%%%%%%%%%%%%%%%
%%%%%%%%%%%%%%%%%%%%%%%%%%%%%%%%%%%%%%%%%%%%%%%%%%%%%%%%%%%%%%%%%%%%%%%%
\section{Results \& discussion}\label{Res}
In order to test the predictions from Section \ref{Theory}, we perform 
extensive Direct Numerical Simulations (DNS) of particle-pair dispersion in 2D turbulence. Owing to the
very high resolutions that are required in order to accurately resolve both the direct and inverse cascades of 2D turbulence, we here
focus on the dispersion in the inverse cascade regime, and will consider the behavior in the direct cascade regime
in a future work.

We integrate the Navier-Stokes equation for the vorticity field
$\omega\equiv{\bm \nabla} \times {\bm u}$
\begin{equation}
\partial_t \omega + {\bm u\cdot} {\bm \nabla} \omega = \nu_p 
\nabla^{2p} \omega - \alpha \omega + f,
\label{dns1}
\end{equation}
in a square box of size $L=2 \pi$ with periodic boundary conditions
using a fully dealiased pseudo-spectral code with second-order
Runge-Kutta time stepping \cite{boffetta00}. We also use a $p=8$ 
hyperviscous dissipation in order to extend the inertial range.
The friction term proportional to $\alpha$ is necessary to avoid
condensation of energy at the largest scale and to reach a statistically
stationary state.  The flow is generated 
by a small scale, $\delta$-correlated in time random forcing $f$,
concentrated at the scale $\ell_f$, which injects energy at a rate
$\varepsilon_I$. This defines the forcing timescale 
$\tau_f=(\ell_f^2/\varepsilon)^{1/3}$ which is used to rescale 
temporal variables.

For each value of $\tau_p$, we inject $M=65536$ particles 
with random initial positions $\bm{x}^p(0)$ and $\bm{v}^p(0)=\bm{0}$. 
Particles are advected according to 
\begin{equation}
\ddot{\bm{x}}^p(t)\equiv\dot{\bm{v}}^p(t)=\frac{1}{\tau_p}
\Big(\bm{u}(\bm{x}^p(t),t)-\bm{v}^p(t)\Big),
\label{dns2}
\end{equation}
for a large-scale time until their distribution in the phase-space 
becomes stationary, after which we collect their trajectories
for a time $T=540 \tau_f$ \cite{boffetta04}. The simulations were performed at a resolution of $1024^2$ with forcing centered on mode 320. The inverse cascade inertial range extended to large scale $L=u_{\rm rms}^3/\epsilon\simeq 120\ell_f$, corresponding to a large scale time $\tau_L=u_{\rm rms}^2/\epsilon\simeq 24\tau_f$.

The statistics of particle separation, both forward and backward in time,
are computed offline from these trajectories by looking, at each time,
at particle pairs which are at the reference separation ${\xi}$. In what follows, the Stokes number is defined via the characteristic forcing timescale ${\rm St}=\tau_p/\tau_f$.

\begin{figure}[ht]
\centering
\centering
\subfloat[]
{\hspace{3mm}\begin{overpic}
[trim = 8mm 12mm 0mm 0mm,scale=0.65,clip,tics=20]{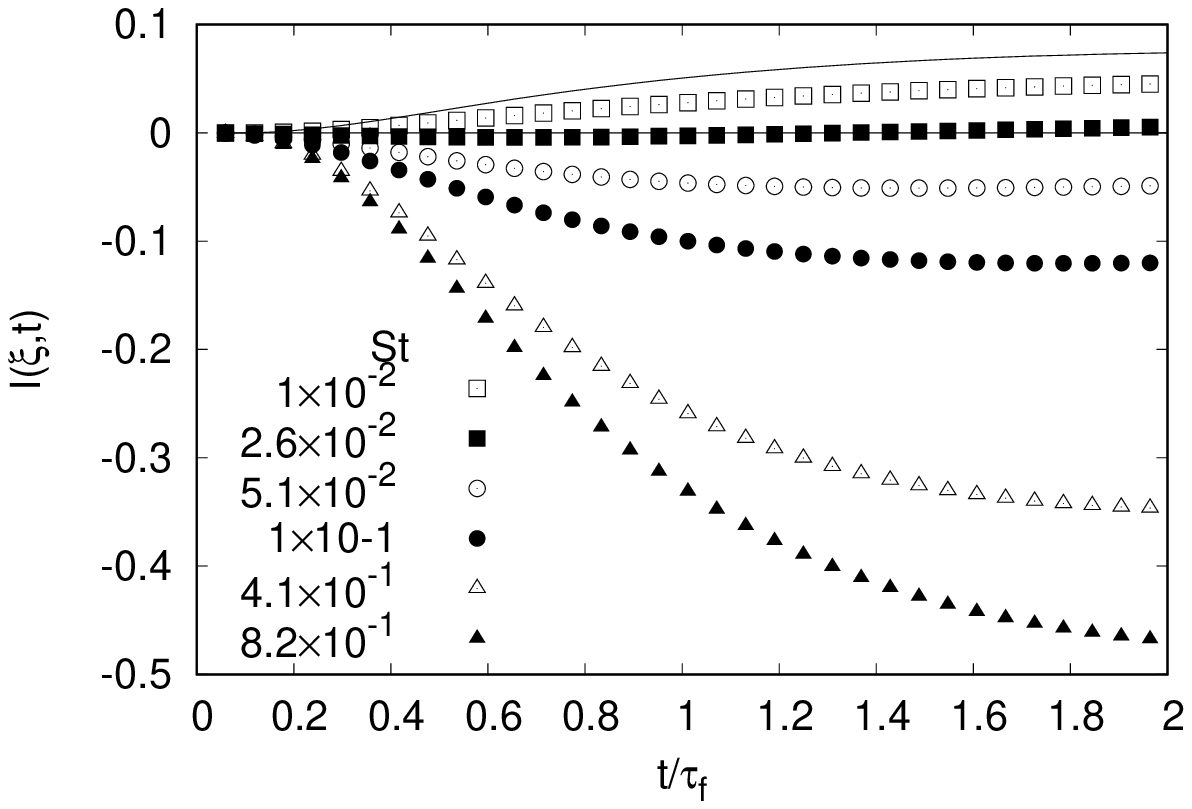}
\put(100,-10){$t/\tau_f$}
\put(-10,55){\rotatebox{90}{$\mathcal{I}(\xi,t)$}}
\end{overpic}}
\subfloat[]
{\hspace{3mm}\begin{overpic}
[trim = 8mm 12mm 0mm 0mm,scale=0.65,clip,tics=20]{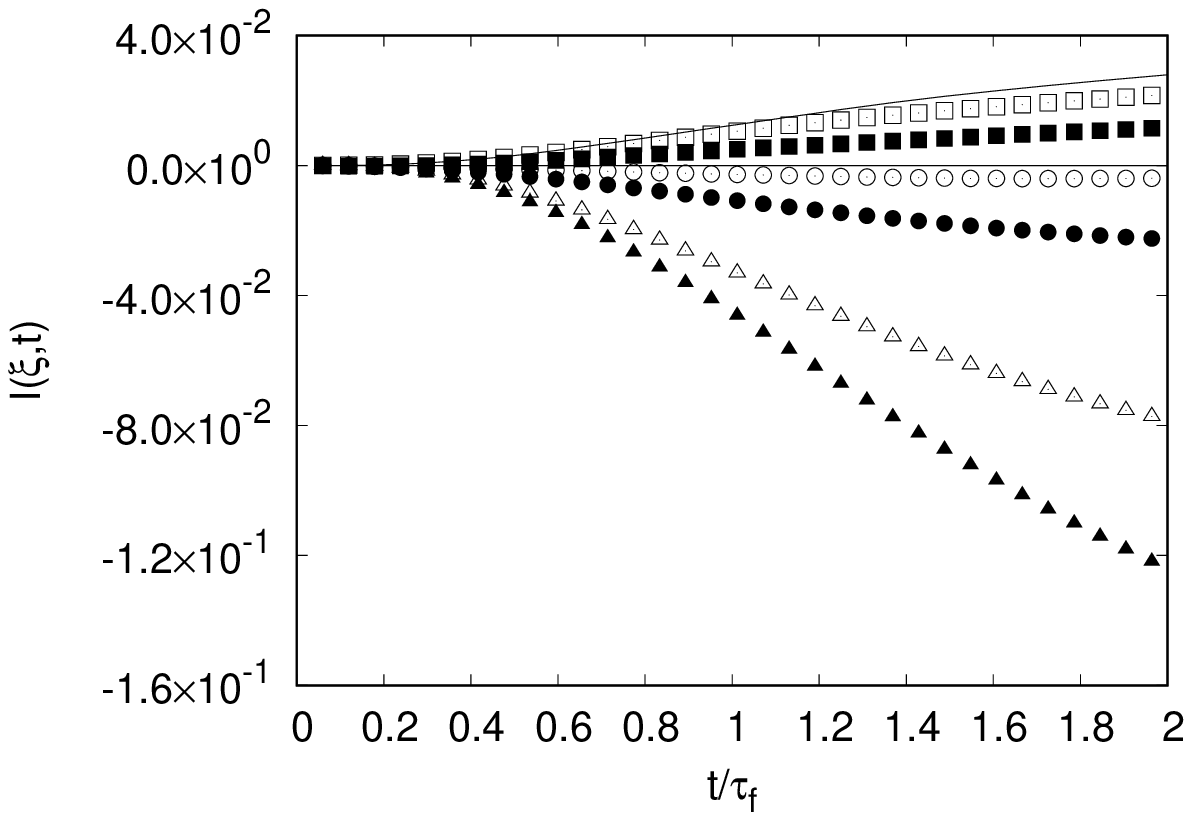}
\put(100,-10){$t/\tau_f$}
\put(-10,55){\rotatebox{90}{$\mathcal{I}(\xi,t)$}}
\end{overpic}}\\
\subfloat[]
{\hspace{3mm}\begin{overpic}
[trim = 8mm 12mm 0mm 0mm,scale=0.65,clip,tics=20]{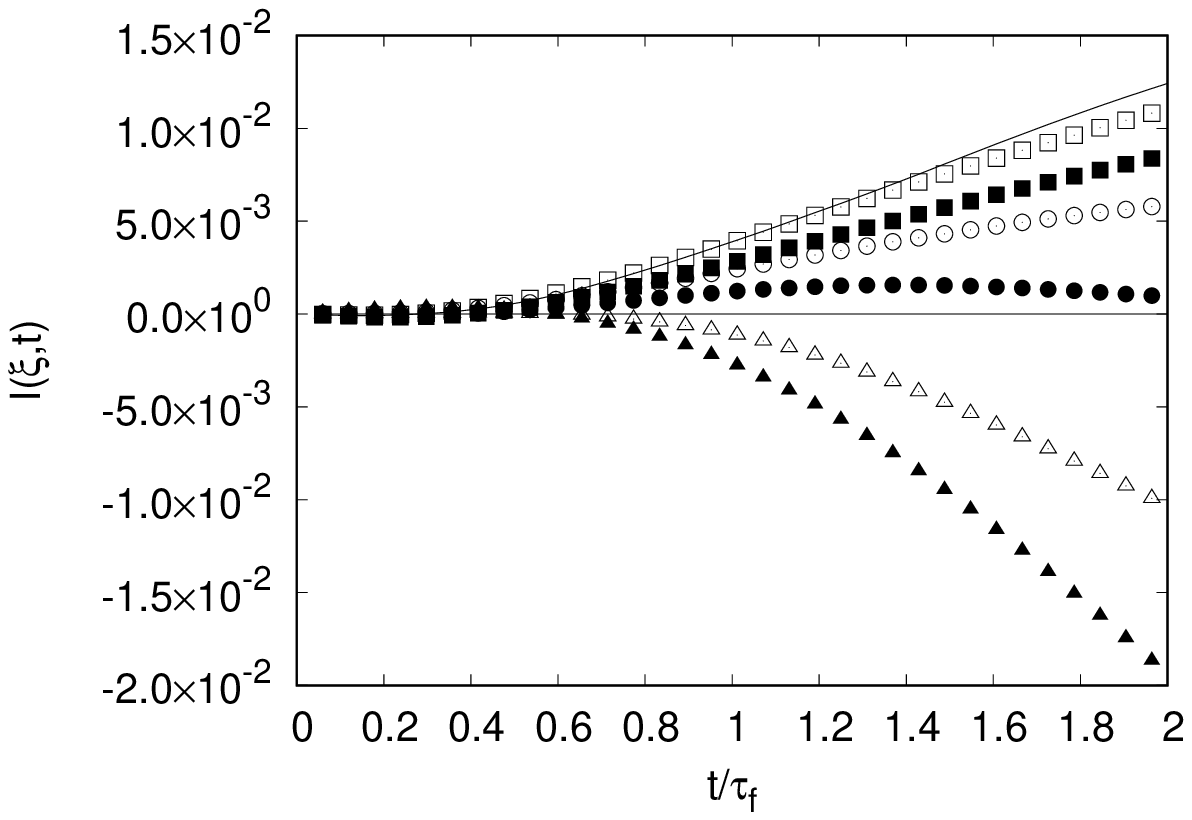}
\put(100,-10){$t/\tau_f$}
\put(-10,55){\rotatebox{90}{$\mathcal{I}(\xi,t)$}}
\end{overpic}}
\subfloat[]
{\hspace{3mm}\begin{overpic}
[trim = 8mm 12mm 0mm 0mm,scale=0.65,clip,tics=20]{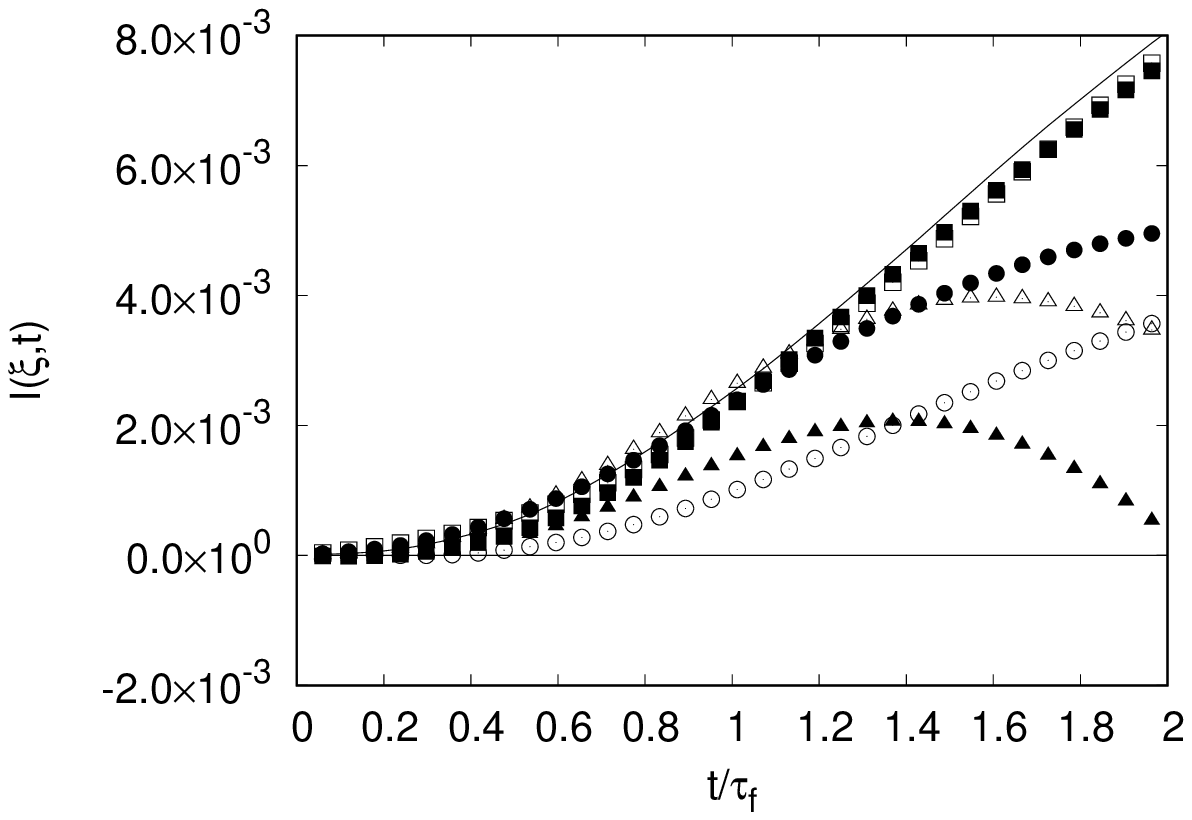}
\put(100,-10){$t/\tau_f$}
\put(-10,55){\rotatebox{90}{$\mathcal{I}(\xi,t)$}}
\end{overpic}}
\caption{DNS results for $\mathcal{I}(\xi,t)$ for various ${\rm St}$ and (a) $\xi/\ell_f=2$, (b) $\xi/\ell_f=5$, (c) $\xi/\ell_f=8$, (d) $\xi/\ell_f=10$. Solid line denotes ${\rm St}=0$ results, and the horizontal line indicates $\mathcal{I}=0$.}
\label{Iplot}
\end{figure}

%
%%%%%%%%%%%%%%%%%%%%%%%%%%%%%%%%%%%%%%%%%%%%%%%%%%%%%%%%%%
%
In Fig.~\ref{Iplot} we consider results for
\begin{align}
\mathcal{I}(\xi,t)\equiv \frac{\Big\langle\|\bm{r}^p(t)\|^2-\|\bm{r}^p(-t)\|^2\Big\rangle_{\xi}}{\Big\langle\|\bm{r}^p(-t)\|^2\Big\rangle_{\xi}},
\end{align}
defined such that $\mathcal{I}(\xi,t)>0$ denotes FIT is faster than BIT dispersion,  $\mathcal{I}(\xi,t)<0$ denotes BIT is faster than FIT dispersion, and  $\mathcal{I}(\xi,t)=0$ denotes reversible dispersion (also, trivially,  $\mathcal{I}({\xi},0)\equiv0$). 

The results confirm the theoretical prediction of \S\ref{Theory}, showing (for a given $\xi$) a transition from $\mathcal{I}<0$ to $\mathcal{I}>0$ as ${\rm St}$ is increased. 
%In some cases, for a fixed $St_\Omega$ and $\xi$, $\mathcal{I}$ changes sign as $t$ is increased. 
%
\FloatBarrier
The results also show that for a given ${\rm St}$, the sign of $\mathcal{I}$ can change as $\xi$ is increased. This is because for fixed ${\rm St}$, as $\xi$ is increased (and therefore ${\rm St}_\xi$ decreased), the NLIM weakens, and at sufficiently large scales, ${\rm St}_\xi$ becomes small enough for the LIM to dominate, giving $\mathcal{I}>0$.

It is interesting to note that the irreversibility inversions occur even when ${\rm St}$ is small. This is not in contradiction to the arguments in \S\ref{Theory} since the regimes ${\rm St}_\xi \ll1$ and ${\rm St}_\xi\geq O(1)$ are merely asymptotically defined, and the latter range is simply intended to denote the range over which the NLIM is expected to operate. We expect that the sign of $\mathcal{I}$ can change even for small ${\rm St}_\xi$ because the fluid energy flux in 2D turbulence is very small. As a result, only a small contribution from the path-history effect is needed to reverse the sign of the particle-pair energy flux, and hence reverse the sign of $\mathcal{I}$. This can also be understood in light of results in \cite{bragg17b} that show that for 3D turbulence, the effect of the path-history mechanism on the odd-order moments of $\bm{w}^p(t)$ is very strong even for ${\rm St}\ll1$, in which regime its effect on the even order moments is small.
\begin{figure}[ht]
\centering
\subfloat[]
{\hspace{3mm}\begin{overpic}
[trim = 8mm 12mm 0mm 0mm,scale=0.65,clip,tics=20]{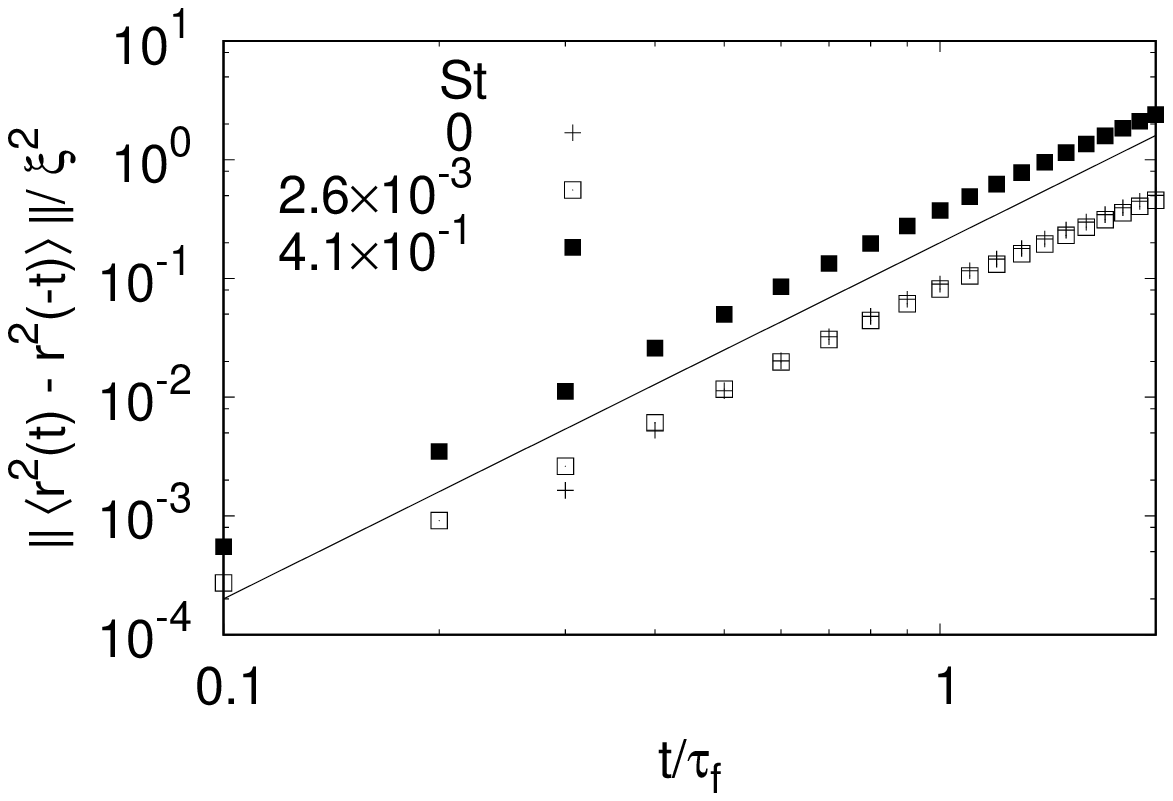}
\put(100,-5){$t/\tau_f$}
\put(-22,10){\rotatebox{90}{\footnotesize{$\xi^{-2}\Big\vert\langle\|\bm{r}^p(t)\|^2-\|\bm{r}^p(-t)\|^2\rangle_{\xi}\Big\vert$}}}
\end{overpic}}
\hspace{4mm}
\subfloat[]
{\hspace{-4mm}\begin{overpic}
[trim = 8mm 12mm 0mm 0mm,scale=0.65,clip,tics=20]{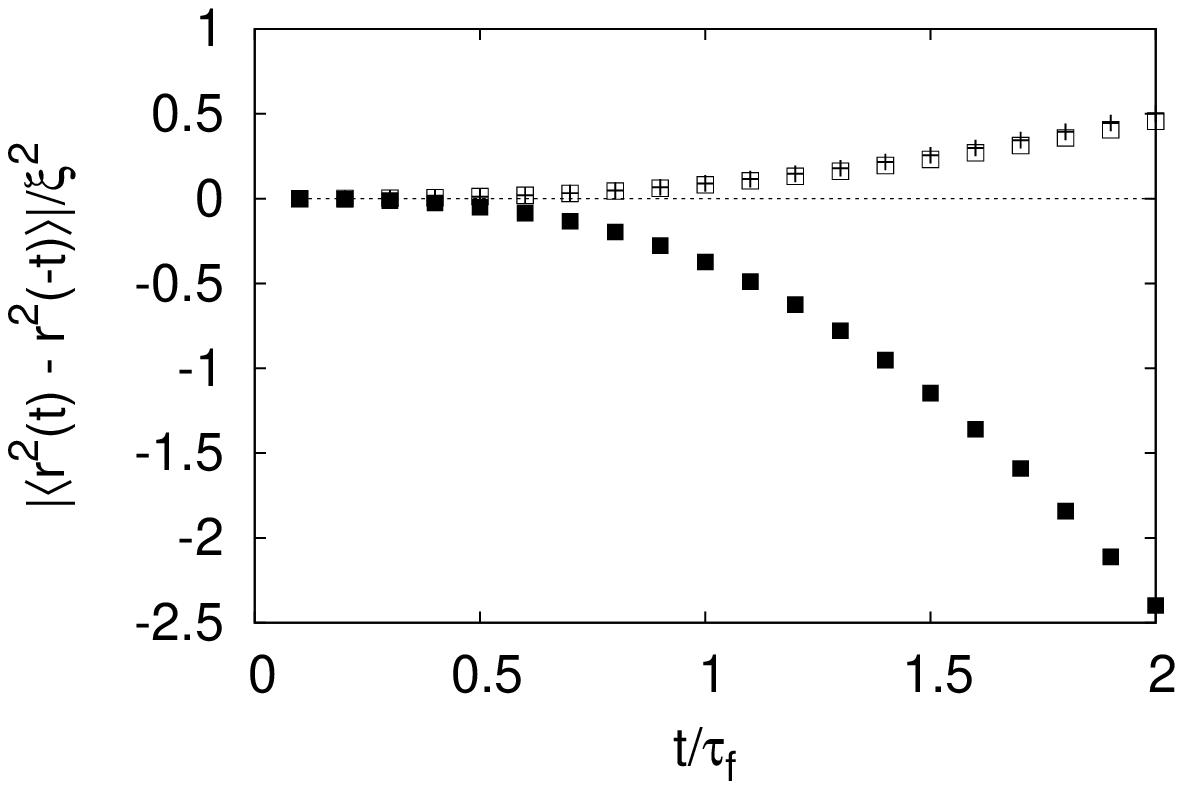}
\put(109,-5){$t/\tau_f$}
\put(-2,15){\rotatebox{90}{\footnotesize{$\xi^{-2}\langle\|\bm{r}^p(t)\|^2-\|\bm{r}^p(-t)\|^2\rangle_{\xi}$}}}
\end{overpic}}
\caption{DNS results for (a) $\xi^{-2}\vert\langle\|\bm{r}^p(t)\|^2-\|\bm{r}^p(-t)\|^2\rangle_{\xi}\vert$ and (b) $\xi^{-2}\langle\|\bm{r}^p(t)\|^2-\|\bm{r}^p(-t)\|^2\rangle_{\xi}$ for various $St$ and $\xi/\ell_f=2$. Whereas plot (a) emphasizes the $t^3$ scaling, plot (b) reveals the change in sign of the short-time behavior of $\langle\|\bm{r}^p(t)\|^2-\|\bm{r}^p(-t)\|^2\rangle_{\xi}$ as ${\rm St}$ is increased, signifying the irreversibility inversion.}
\label{t3plot}
\end{figure}
\FloatBarrier
In Figure~\ref{t3plot}(a) we plot $\vert\langle\|\bm{r}^p(t)\|^2-\|\bm{r}^p(-t)\|^2\rangle_{\xi}\vert$ in order to test the prediction of \S\ref{Theory} that for small $t$, $\langle\|\bm{r}^p(t)\|^2-\|\bm{r}^p(-t)\|^2\rangle_{\xi}\propto t^3$. The results confirm the short-time $t^3$ growth of $\langle\|\bm{r}^p(t)\|^2-\|\bm{r}^p(-t)\|^2\rangle_{\xi}$ well, for ${\rm St}\geq0$. In Figure~\ref{t3plot}(b) we highlight the change in sign of $\langle\|\bm{r}^p(t)\|^2-\|\bm{r}^p(-t)\|^2\rangle_{\xi}$ in the short time $t^3$ regime, as ${\rm St}$ is increased.

\begin{figure}[ht]
\centering
\subfloat[]
{\hspace{-6mm}\begin{overpic}
[trim = 6mm 10mm 0mm 0mm,scale=0.6,clip,tics=20]{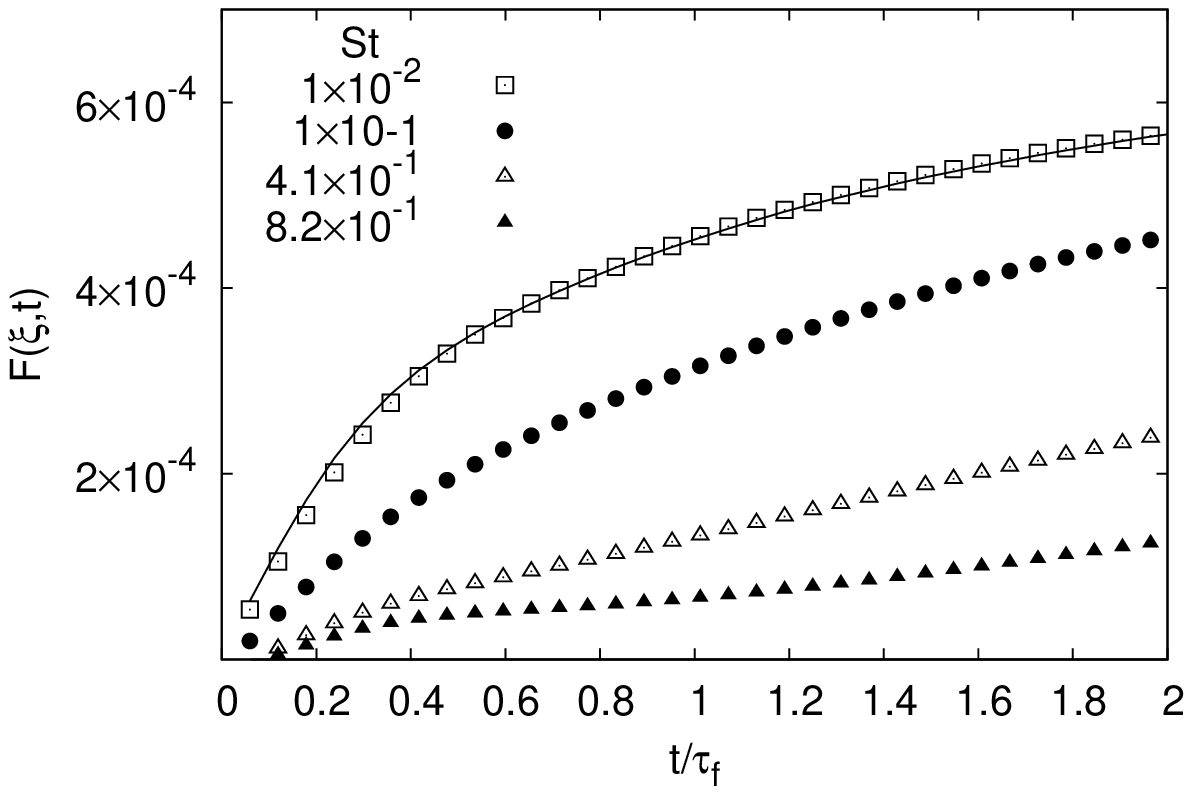}
\put(102,-7){$t/\tau_f$}
\put(-10,50){\rotatebox{90}{$\mathcal{J}(\xi,t)$}}
\end{overpic}}
\hspace{9mm}
\subfloat[]
{\hspace{-6mm}\begin{overpic}
[trim = 6mm 10mm 0mm 0mm,scale=0.6,clip,tics=20]{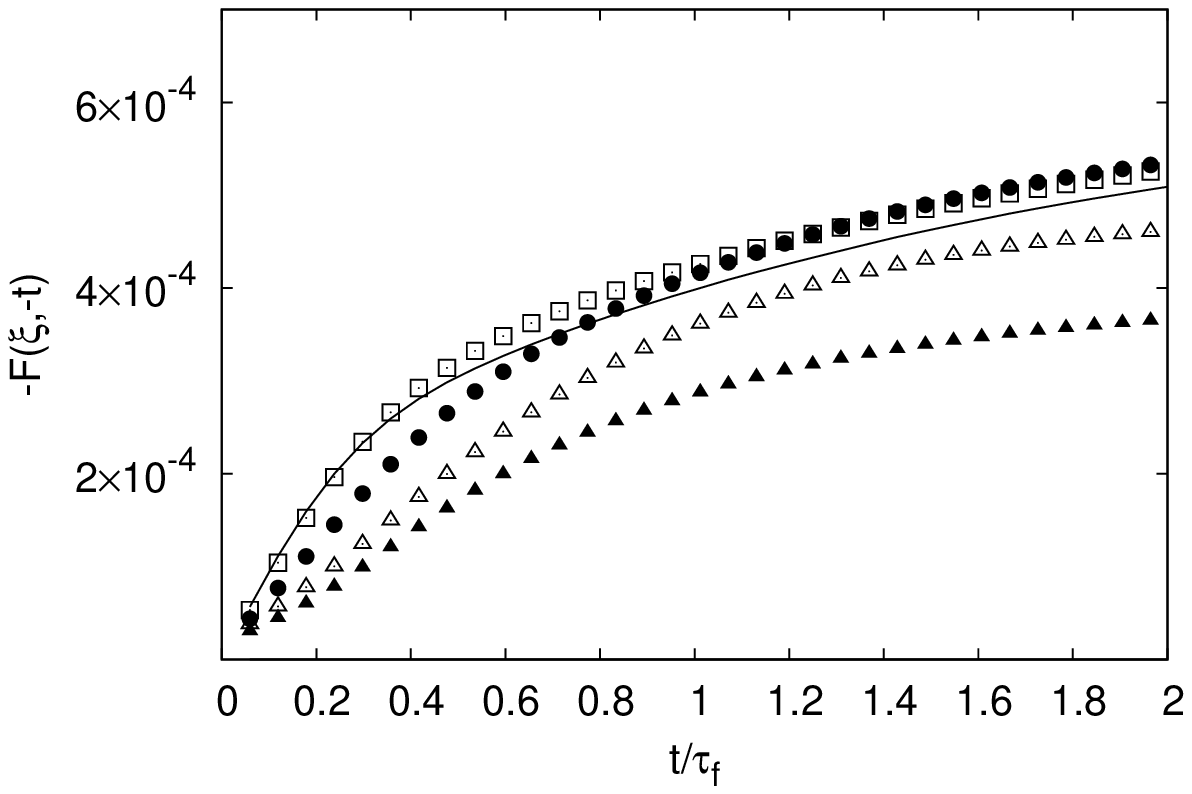}
\put(102,-7){$t/\tau_f$}
\put(-10,45){\rotatebox{90}{$-\mathcal{J}(\xi,-t)$}}
\end{overpic}}
\caption{DNS results for (a) $\mathcal{J}(\xi,t)$ and (b) $\mathcal{J}(\xi,-t)$ for various ${\rm St}$ and $\xi/\ell_f=2$.}
\label{Lag_skew}
\end{figure}

We now use the DNS data to test our theoretical explanations for the dispersion irreversibility. First, we assumed that $\langle\|\bm{r}^p(t)\|^2\rangle_{\xi}$ is dominated by the behavior of particle-pairs that are separating, and therefore have $w^p_\parallel>0$, whereas $\langle\|\bm{r}^p(-t)\|^2\rangle_{\xi}$ is dominated by the behavior of particle-pairs that are approaching, and therefore have $w^p_\parallel<0$ \cite{bragg16}. One way to test this assumption is to compute from the DNS the quantities
\begin{align}
\mathcal{J}(\xi,t)\equiv \Big\langle \Big[w^p_\parallel(t)\Big]^3 \Big\rangle_{r^p(0)=\xi},\quad \mathcal{J}(\xi,-t)\equiv \Big\langle \Big[w^p_\parallel(-t)\Big]^3 \Big\rangle_{r^p(0)=\xi}.
\end{align}
The results in Fig.~\ref{Lag_skew} clearly validate our argument, showing $\mathcal{J}(\xi,t)>0$ and $\mathcal{J}(\xi,-t)<0$.
\FloatBarrier
Second, we argued that at any given separation $r$, particle-pairs that are moving together should do so with relative velocities whose magnitudes are characteristically different from those of particles that are moving apart. In other words, the PDF of $w^p_\parallel$ should be skewed, both because of the presence of dynamical fluxes in the turbulent velocity field, and because of the path-history effect. Our arguments predict that in 2D turbulence, when ${\rm St}_\xi\ll1$, the skewness should be positive, but when ${\rm St}_\xi$ is large enough for the non-local inertial particle dynamics to dominate the flux, the skewness should become negative. The quantity $\mathcal{J}$ is not appropriate for testing this part of the argument; since the particles disperse at different rates FIT and BIT, then for a given $t$, $\mathcal{J}(\xi,t)$ and $\mathcal{J}(\xi,-t)$ will be associated with the particle relative velocities at different separations. Instead, the appropriate quantity to test this argument is
\begin{align}
\mathcal{S}^p_{\parallel}(r)\equiv \Big\langle \Big[w^p_\parallel(t)\Big]^3 \Big\rangle_{r}\Bigg/\Big\langle \Big[w^p_\parallel(t)\Big]^{2} \Big\rangle_{r}^{3/2},
\end{align}
i.e. the skewness of the PDF of $w^p_\parallel$ at fixed separation $r$.
\begin{figure}[ht]
\centering
{\hspace{-6mm}\begin{overpic}
[trim = 7mm 10mm 0mm 0mm,scale=0.75,clip,tics=20]{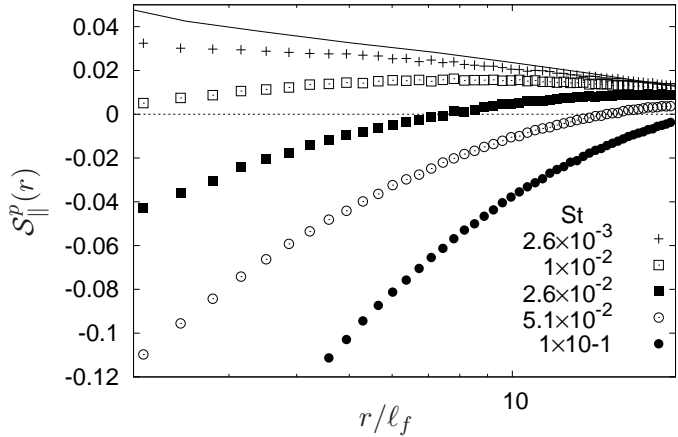}
\put(120,-2){$r/\ell_f$}
\put(-12,70){\rotatebox{90}{$\mathcal{S}^p_{\parallel}(r)$}}
\end{overpic}}
\caption{DNS results for $\mathcal{S}^p_{\parallel}(r)$ for various ${\rm St}$. The continuous line is the skewness of the fluid velocity (i.e. ${\rm St}=0$), given for reference. Note the change in sign of $\mathcal{S}^p_{\parallel}(r)$ as ${\rm St}$ is increased.}
\label{Skew}
\end{figure}
\FloatBarrier
The results in Figure~\ref{Skew} confirm the predictions; for fluid particles  $\mathcal{S}^p_{\parallel}>0\,\forall r$, whereas $\mathcal{S}^p_{\parallel}$ can be both positive or negative for inertial particles, depending upon $r$ and ${\rm St}$. Also in agreement with the theoretical predictions, for a given ${\rm St}$, $\mathcal{S}^p_{\parallel}$ can change sign as $r$ is varied. This is due to the variation in ${\rm St}_r$ with $r$: At small enough $r$, ${\rm St}_r$ can be large enough for the non-local inertial effects to dominate, yielding $\mathcal{S}^p_{\parallel}<0$. However, as $r$ is increased, ${\rm St}_r$ decreases, and when ${\rm St}_r$ becomes sufficiently small the local turbulence dominates the inertial particle behavior, and $\mathcal{S}^p_{\parallel}>0$.

The results in Figure~\ref{Skew} are of significant interest even beyond the problem of dispersion irreversibility. In particular, they show that even though the fluid exhibits an inverse energy flux in 2D, particles transported by such a flow may in fact exhibit a downscale/direct energy flux, depending on their inertia. This non-trivial behavior is yet another manifestation of the complexity and subtlty of inertial particle dynamics in turbulent flows.

It is important to emphasize that the qualitative explanations given in \S\ref{Theory} connect the irreversibility of the dispersion to the asymmetry of the PDF of $w^p_\parallel$ \emph{for any time} in the dispersion process. However, at present we are only able to demonstrate this analytically in the limit $t\to 0$, through the analysis in \S\ref{Theory}. An important point for future work is to demonstrate this dependence analytically for arbitrary $t$, which is a very challenging task.  

We close this section with a comment on the relationship between the local
irreversibility mechanism and the dynamical cascade processes in operation in
2D turbulence. According to the theoretical explanations in 
\cite{bragg16} and those in \S\ref{Theory}, the local irreversibility
mechanism is connected to the sign of the fluid energy flux, irrespective of
the underlying dynamics responsible for this. For example, in the present case
of 2D turbulence, our explanations predict that FIT is faster than BIT
dispersion for ${\rm St}=0$, irrespective of whether the particle separation lies in
the regime of the inverse energy cascade or the direct enstrophy cascade. All
that matters for this prediction is that the flux is positive. Our explanation
is therefore somewhat different to the explanation proposed in 
\cite{faber09}, who connected the irreversibility of fluid
particle-pair dispersion in 2D turbulence with the dynamics of the inverse
energy cascade itself. However, we emphasize that through the local
irreversibility mechanism, fluid particle-pair dispersion would be irreversible
even in kinematically constructed flow fields, provided only that the PDF of
$\Delta u_\parallel$ is asymmetric and that the Lagrangian timescales of the
flow are finite.
\section{Conclusions}\label{Conc}

In this paper we have supplemented a qualitative argument presented in a recent
paper with a new quantitative prediction for the irreversibility of inertial
particle dispersion in 2D turbulence. Using DNS data we have confirmed the
predictions that in 2D turbulence, the forward dispersion of particle-pairs is
faster than the backward dispersion, until the particle inertia passes a
certain threshold, and then the backward dispersion becomes faster than the
forward dispersion. The confirmation of the prediction lends strong support to
our arguments that the irreversibility of inertial particle dispersion in
turbulence is governed by two completely distinct physical mechanisms, whose
relative influence depends upon the inertia of the particles. More generally,
the results are also of interest since they reveal that in turbulence flows
with an inverse energy cascade, inertial particles may exhibit a downscale flux
of kinetic energy because of their non-local in-time dynamics. These results
could therefore be significant in understanding and modeling the motion of
inertial particles in certain geophysical and astrophysical flows that exhibit
quasi-2D dynamics.

\bibliography{refs_co12}

\end{document}